\documentstyle[eqsecnum,aps,epsfig]{revtex}
\begin{document}
\draft
\preprint{Draft Nb. 1}
\title{
Simulation accuracy of long range muon propagation in medium:\\ 
analysis of error sources  
}
\author{Edgar V. Bugaev, Igor A. Sokalski, Sergey I. Klimushin}
\address{
Institute for Nuclear Research, Russian Academy of Science, 
60th October Anniversary prospect 7a, Moscow 117312, Russia
}
\date{\today}
\maketitle
\begin{abstract}
Knowledge of atmospheric muon flux intensity at large depths is extremely 
important for neutrino telescopes located deep under ground, water or ice.
One of the methods to transform muon sea-level spectrum into depth one is 
to apply Monte Carlo technique which directly takes into account 
stochastical nature of energy loss. In order to decrease computation time 
down to acceptable level one has to use simplifications resulting in 
systematic errors which in some cases may distort result essentially. Here 
in this paper we present our analysis for dependence of computed 
depth muon flux upon the most important parameters of muon transport Monte 
Carlo algorithm which was done with the MUM (MUons+Medium) code. 
Contribution of different simplifications to the resulting error is 
considered, ranked and compared with uncertainties which come from 
parametrization accuracy both for sea-level muon spectrum and for muon 
cross sections.  
\end{abstract}
\pacs{PACS number(s): 13.85Tp, 96.40.Tv, 02.70Lq}

\widetext

\section{INTRODUCTION}
\label{sec:int1}

In order to obtain muon spectrum and value for integral muon flux at large 
depths where existing and planned neutrino telescopes are located (one to 
several kilometers of water equivalent) one has to perform several steps 
starting with interactions of primary cosmic rays in atmosphere. The last 
step of this procedure is transportation of muons from sea level down to 
detector location. It can be done both with analytical or semianalytical 
technique and Monte Carlo (MC) simulation 
which directly takes into account that muon energy loss represents an 
essentially stochastic process. MC codes for  muon transportation have to 
use some simplifications to optimize equilibrium  between reasonable 
computation time, desirable accuracy and neccessary statistics. 

The usual approximation when calculating the muon propagation through a
medium is the following: muon interactions with comparatively large energy
transfers, i.e., when fraction of energy lost $v=\Delta E/E$ exceeds some 
value $v_{cut} \sim$ 10$^{-3}$--10$^{-2}$, are taken into account by direct
simulation of $v \ge v_{cut}$  for each single interaction according to 
shape of differential cross sections $d\sigma(E,v)/dv$ (these interactions 
lead to ``stochastic'' energy loss or SEL) while the part of interaction 
with relatively small $v$ is treated by the approximate concept of 
``continuum'' energy loss (CEL), i.e., using the function 
$[dE(E)/dx]_{CEL}$ which is obtained from formula
\begin{eqnarray}
\left[
\frac{dE}{dx}(E)
\right]
_{CEL} &\! = & \frac{N_{A}}{A}E
\left(\:
\int\limits_{v_{min}^{\gamma}}^{v_{cut}}\frac{d\sigma^{\gamma}(E,v)}{dv}v\,dv
\right) + 
\frac{N_{A}}{A}E
\left(\:
\int\limits_{v_{min}^{e^{+}e^{-}}}^{v_{cut}}\frac{d\sigma^{e^{+}e^{-}}(E,v)}{dv}v\,dv
\right) \nonumber\\
&\!+&
\frac{N_{A}}{A}E
\left(\:
\int\limits_{v_{min}^{pn}}^{v_{cut}}\frac{d\sigma^{pn}(E,v)}{dv}v\,dv
\right) + 
\left[
\frac{dE}{dx}(E)
\right]_{B-B} -
\frac{N_{A}}{A}E
\left(\:
\int\limits_{v_{cut}}^{v_{max}^{\delta}}\frac{d\sigma^{\delta}(E,v)}{dv}v\,dv
\right) .
\label{CL}
\end{eqnarray} 
Here $N_{A}$ is the Avogadro number; $A$ is the atomic weight; indexes
$\gamma$, $e^{+}e^{-}$, $pn$ and $\delta$ correspond to bremsstrahlung,
direct $e^{+}e^{-}$-pair production, photonuclear interaction and knock-on 
electron production, respectively; $v_{min}$ and $v_{max}$ are the minimum
and the maximum kinematically allowed fraction  of  energy lost for 
corresponding process; term $[dE(E)/dx]_{B-B}$ represents Bethe-Bloch 
formula with correction for density effect. 
Notice that actually one has to compute
CEL, as well as all total cross sections separately for each kind of atoms 
given material consists of and then add them to each other with different 
weights but here and below in the text we omit this detail and give only 
general expressions. One is forced to decompose energy loss because 
simulation of all interactions with $v \ge v_{min}$ would result in 
infinite computation time due to steep dependence of muon cross sections on
$v$ (they decrease with $v$ at least as $d\sigma(E,v)/dv \propto v^{-1}$ 
and for some processes are not finite at $v\to$ 0). Number of interactions 
to be simulated per unit of muon path grows, roughly, as 
$N_{int} \propto v_{cut}^{-1}$ along with computation time. On the other 
hand,  setting $v_{cut}$ to large value may affect simulation accuracy. 
Thus, the question is {\it how large value of $v_{cut}$ may be chosen to 
keep result within desirable accuracy}?
 
The second problem can be formulated as follows:  {\it if it is necessary 
to include ionization in SEL, at all}? Small energy transfers strongly 
dominate at knock-on electron production 
($d\sigma^{\delta}(E,v)/dv \propto v^{-2}$), so this process is almost 
non-stochastic and it seems to be reasonable to omit the last term in 
expression for CEL (Eq.~(\ref{CL})) and exclude knock-on electrons from 
simulation procedure  when simulating SEL, i.e., to treat ionization 
completely as ``continuous'' process which saves computation time 
noticeable. How much does it affect the result of simulation?

Influence of these factors on simulated result was discussed in
literature (see, e.g., Refs.~\cite{N94,lipari,music1,lagutin1}) but, in our
opinion, more detailed analysis on which further improvement of muon 
propagation MC algorithms could be based is still lacking.
In presented work we undertook an attempt to investigate in
details with the MUM (MUons+Medium) muon propagation code (Ref.~\cite{MUM})
the influence of $v_{cut}$ and model of ionization energy loss
upon result of MC simulation. Resulting errors are compared with
uncertainties which come from parametrization accuracy both for sea-level 
muon spectrum and for muon cross sections. Our analysis allows to rank 
different uncertainties according to their importance, expose ones to 
which special attention should be put and choose the most 
adequate setting of MC parameters at simulation for this or that purpose.  
 
We briefly describe the main features of the MUM code which was a basic 
instrument for reported investigation and give a short review of principal
variables have been studied in Sec.~\ref{sec:method}.
Sec.~\ref{sec:results} presents results of simulation for survival 
probabilities and muon flux intensities with different settings of 
simulation paremeters. In Sec.~\ref{sec:discussion} we analyse results of 
simulations presented in Sec.~\ref{sec:results}, compare them with other 
published results and give some basic inferences. 
Sec.~\ref{sec:conclusions} gives general conclusions.

\section{Method}
\label{sec:method}

Description of the MUM code which was used to obtain presented results has
been published in Ref.~\cite{MUM}. Here we only dwell briefly on those 
features of the code which seem to be important in the frame of discussed
problem:

\begin{itemize}
\item[-] the most recent results on parametrizations for muon
cross sections are used in the code; 
\item[-] we tried to decrease the ``methodical'' part of systematic error 
due to interpolation, numerical integration, etc., down to as low level as
possible, especial attention was put on simulation procedures for free path 
and fraction of energy lost;
\item[-] the most important parameters, like value of $v_{cut}$, model for
ionization loss, kind of medium, parametrizations for muon cross sections 
are changeable and represent input parameters for initiation routine; 
\end{itemize} 

The accuracy of simulation  algorithm used in MUM was shown in 
Ref.~\cite{MUM} to be high enough to perform systematically significant 
analysis which is presented below.

In order to study how different factors influence upon result of simulation
we performed several sets of simulations both for propagation of 
monoenergetic
muon beam and muons sampled by real sea-level spectrum (in the later case we
limited ourselves by simulation  only vertical muons) through pure water
down to depths from $D =$ 1 km to $D =$ 40 km. 
Of course, depths of more than
several kilometers of water for vertical muons do not concern any real 
detector but simulations for larger depths allow us to study general 
appropriatenesses for large muon ranges which correspond, for instance, to 
nearly horizontal muons. Several runs were done for standard rock ($A=$ 22,
$Z=$ 11, $\rho=$ 2.65 g cm$^{-3}$), as well. Muons whose energy decreased 
down to $E$ = 0.16 GeV (the Cherenkov threshold for muon in water) were
considered as stopped ones. We tested different settings of parameters 
which were as follows.

\begin{itemize}
\item[(a)] $v_{cut}$, which changed within a range of 
$v_{cut}=$ 10$^{-4}$ to
$v_{cut}=$ 0.2. Actually, ``inner''  accuracy of the MUM code becomes
somewhat worse at $v_{cut} >$ 5$\times$10$^{-2}$, especially if 
fluctuations in ionization are not simulated (Ref.~\cite{MUM}). So, results
for $v_{cut}$ = 0.1 for $v_{cut}$ = 0.2 are presented here only to
illustrate some general qualitative appropriatenesses.
\item[(b)] Model for ionization loss which was treated both as completely
``continuous'' and ``stochastic'' with simulation of energy lost if
knock-on electron energy $\Delta E \ge v_{cut}\,E$.
\item[(c)] Parametrization for vertical sea-level conventional 
atmospheric muon spectrum.
Two spectra were tested, namely one proposed in Ref.~\cite{bks1} (basic):
\begin{equation}
\frac{dN}{dE}=\frac{0.175\:E^{-2.72}}{cm^{2}\:s\:sr\:GeV}
\left(
\frac{1}{\displaystyle
  1+\frac{E}{\displaystyle
   103\:GeV}}+
\frac{0.037}{\displaystyle
  1+\frac{E}{\displaystyle
   810\:GeV}}
\right),
\label{bknsspec}
\end{equation}
and widely used Gaisser spectrum (Ref.~\cite{gaisser}):
\begin{equation}
\frac{dN}{dE}=\frac{0.14\:E^{-2.7}}{cm^{2}\:s\:sr\:GeV}
\left(
\frac{1}{\displaystyle
  1+\frac{E}{\displaystyle
   104.6\:GeV}}+
\frac{0.054}{\displaystyle
  1+\frac{E}{\displaystyle
   772.7\:GeV}}
\right).
\label{gaisspec}
\end{equation}
\item[(d)] Parametrization for total cross section for absorbtion of a real
photon of energy $\nu=s/2m_{N}=vE$ by a nucleon at photonuclear interaction
which was treated both according to Bezrukov-Bugaev parametrization 
proposed in Ref~\cite{bb} (basic):
\begin{equation}
\sigma_{\gamma N}=[114.3+1.647\:\ln^{2}(0.0213\:\nu)] \; \mu b
\label{bbpn}
\end{equation}
and ZEUS parametrization (Ref.~\cite{ZEUS}):
\begin{equation}
\sigma_{\gamma N} = (\:63.5\: s^{0.097} + 145\:s^{-0.5}\:) \; \mu b 
\label{zeuspn}
\end{equation}
($\nu$ and $s$ in GeV and GeV$^{2}$, correspondingly).
\item[(e)] A factor $k_{\sigma}$ which all muon cross sections were 
multiplied by to test influence of uncertainties in cross sections 
parametrization (and, concequently, in energy loss) upon result. We used 
$k_{\sigma}$ = 1.0 as a basic value but in some cases set also 
$k_{\sigma}$ = 0.99 and $k_{\sigma}$ = 1.01, which corresponds to decrease 
and increase of total energy loss by 1\%, 
respectively. Note that it is an ``optimistic'' evaluation, the real 
accuracy of existing parametrization for muon cross sections is worse (see 
Refs.~\cite{kp,rhode1}).
\end{itemize} 

For each run we fixed the muon spectra at final and several interim depths.
The differences between obtained spectra were a point of investigation.

\section{Results}
\label{sec:results}

\subsection{Propagation of monoenergetic muon beams} 
\label{sec:mono}

\begin{figure}
\hspace{1.1cm}
\mbox{\epsfig{file=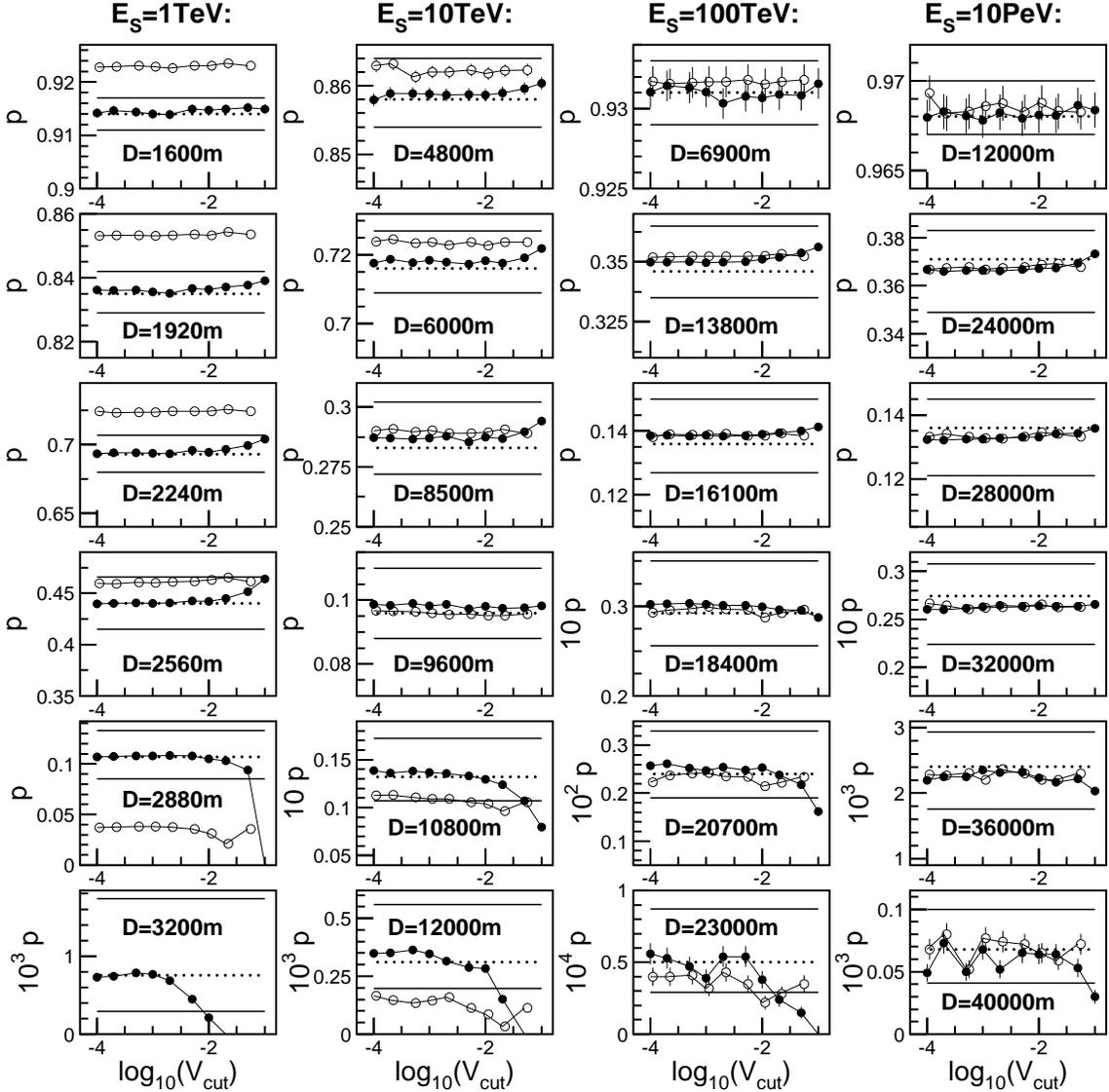,width=15.1cm}}
\protect\caption{
Survival probabilities $p=N_{D}/N_{s}$ (where $N_{s}$ = 10$^{6}$ is initial
number of muons and $N_{D}$ is number of muons which have survived after
propagation down to depth $D$ in pure water) vs. $v_{cut}$. Values of $p$ 
were obtained as a result of MC simulation for monoenergetic muon beams 
with initial energies $E_{s}$ = 1 Tev (1st column of plots), 10 TeV (2nd 
column), 100 TeV (3rd column) and 10 PeV (4th column). Each column contains
six plots which correspond to six depths $D$ (which differs for different 
$E_{s}$). Closed circles represent survival probabilities which were 
simulated with ionization energy loss included in SEL along with other 
types of muon interactions. Open circles correspond to computation with 
completely ``continuous'' ionization. Two horizontal solid lines on each 
plot show the value for survival probability computed with all muon cross 
sections multiplied by a factor $k_{\sigma}$ = 1.01 (lower line) and 
$k_{\sigma}$ = 0.99 (upper line) for $v_{cut}$ = 10$^{-4}$. Horizontal 
dotted lines correspond to $v_{cut}$ = 10$^{-4}$ and cross section for 
absorbtion of a real photon at photonuclear interaction parametrized 
according to recent ZEUS data (Ref. \protect\cite{ZEUS}, 
Eq.~(\protect\ref{zeuspn})) instead of parametrization proposed by L.B. 
Bezrukov and E.V. Bugaev (Ref. \protect\cite{bb}, Eq.~(\protect\ref{bbpn}))
which is basic in the MUM code. Note different scales at Y-axis.
}
\label{fig1}
\end{figure}

At the first set of simulations we propagated monoenergetic muon beams of
4 fixed initial energies $E_{s}$ = 1 TeV, 10 TeV, 100 TeV and 10 PeV down
to depths $D$ = 3.2 km, 12 km, 23 km and 40 km, respectively, through pure 
water. 
In each case propagation of 10$^{6}$ muons was simulated.   
Fig.~\ref{fig1} shows resulting survival probabilities $p=N_{D}/N_{s}$ 
(where $N_{s}$ = 10$^{6}$ is initial number of muons and $N_{D}$ is number 
of muons which have survived after propagation down to depth $D$) vs. 
$v_{cut}$ for final and five interim depths. Two curves are given on each 
plot for two models of ionization. Also results for 
$k_{\sigma}$ = 1.00 $\pm$ 0.01 and for parametrization (\ref{zeuspn}) are 
presented as simulated with the most accurate value $v_{cut}$ = 10$^{-4}$.

The following appropriatenesses are visible.
\begin{itemize}
\item[(a)] In most cases with the exception of 
some plots of the lower row and 
the left column in Fig.~\ref{fig1} (which corresponds to low survival 
probabilities and low muon initial energies, respectively) uncertainty in 
our knowlege of muon cross sections gives the principal effect which 
essentially exceeds ones from other tested parameters.
\item[(b)] The difference between survival probabilities for two models of 
ionization is the less appreciable the larger muon energy is. It is quite 
understandable because at muon energies $E<$ 1 TeV ionization represents 
the great bulk of total energy loss, and vice versa, it becomes minor at 
$E>$ 1 TeV. Thus, contribution which is given by ionization at higher
energies is small and, the more, its fluctuations do not play an important
role. For muons with initial energies $E \gg$ 1 TeV fluctuations in
ionization become important only at very last part of muon path and 
``are not in time'' to produce some noticeable effect.
\item[(c)] Generally, parametrizations (\ref{bbpn}) and (\ref{zeuspn}) do 
not show a noticeable difference in survival probabilities, in most cases 
it is within statistical error or exceeds it only slightly.
\item[(d)] Increase of $v_{cut}$ gives effect of both signs in survival 
probabilities: function $p(v_{cut})$ grows at the beginning of muon path
and falls at the last part. The same ``both-sign'' dependencies are observed
for ionization model. 
\item[(e)] For $v_{cut} \le$ 0.02 -- 0.05 there is almost no dependence of
survival probability on $v_{cut}$ with the exception of very last part
of muon path where survival probability becomes small. Generally,
dependence $p(v_{cut})$ is the less strong the larger initial muon
energy is.
\end{itemize}

\begin{figure}
\hspace{4.2cm}
\mbox{\epsfig{file=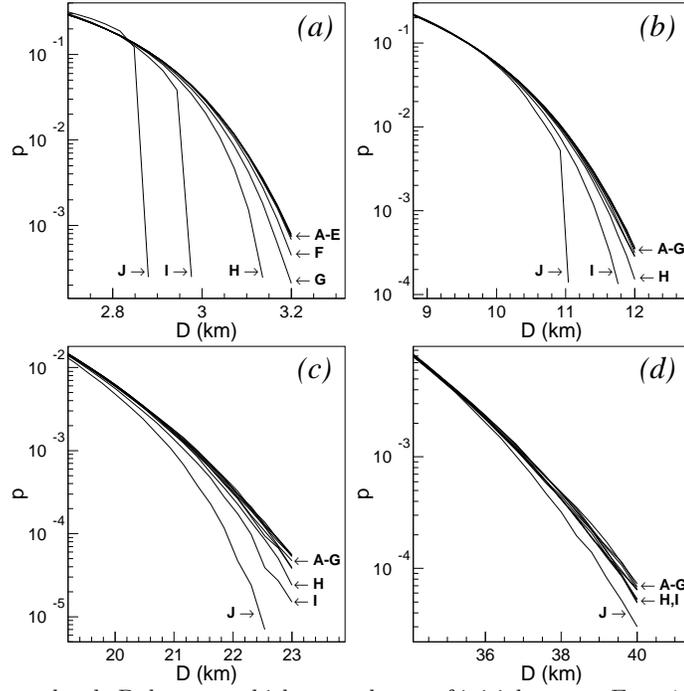,width=9.1cm}}
\caption{Survival probability $p$ vs. depth $D$ down to which muon beam
of initial energy $E_{s}$ = 1 TeV (a), 10 TeV (b), 100 TeV (c) and
10 PeV (d) propagates through pure water. On each plot 10 lettered curves 
which correspond to different values of $v_{cut}$ are shown. Meaning
of letters is as follows: 
A - $v_{cut}$ = \mbox{10$^{-4}$}, 
B - $v_{cut}$ = \mbox{2$\times$10$^{-4}$}, 
C - $v_{cut}$ = \mbox{5$\times$10$^{-4}$}, 
D - $v_{cut}$ = \mbox{10$^{-3}$}, 
E - $v_{cut}$ = \mbox{2$\times$10$^{-3}$}, 
F - $v_{cut}$ = \mbox{5$\times$10$^{-3}$}, 
G - $v_{cut}$ = \mbox{10$^{-2}$}, 
H - $v_{cut}$ = \mbox{2$\times$10$^{-2}$}, 
I - $v_{cut}$ = \mbox{5$\times$10$^{-2}$}, 
J - $v_{cut}$ = \mbox{10$^{-1}$}. 
This figure displays results were obtained by simulation with 
ionization loss included in SEL. Statistical errors which cause some
unsmoothness of curves at small $p$ are not shown. Dependence $p$ upon
$v_{cut}$ becomes noticeable only at the last $\approx$ 1/8 of muon beam 
path where the majority of muons (from 90\% 
to 99.5\%, 
depending upon $E_{s}$) has been stopped. 
}
\label{fig2}
\end{figure}

\begin{figure}
\hspace{4.2cm}
\mbox{\epsfig{file=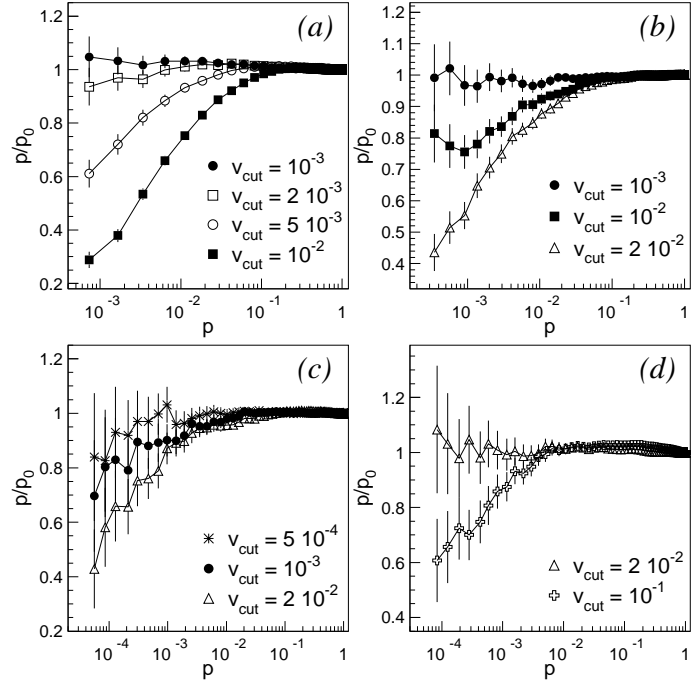,width=9.1cm}}
\caption{Relation $p/p_{0}$ vs. $p$. $p$ is survival probability for muon
beam of initial energy $E_{s}$ = 1 TeV (a), 10 TeV (b), 100 TeV (c) and 
10 PeV (d) at propagation through pure water with ionization included in 
SEL as 
simulated for different values of $v_{cut}$; $p_{0}$ is survival 
probability simulated under the same conditions for $v_{cut}$ = 10$^{-4}$. 
Difference in $p/p_{0}$ becomes noticeable only at small values of $p$, 
i.e., at the last part of muon beam path.
}
\label{fig3}
\end{figure}

The last item is illustrated complementary by Fig.~\ref{fig2} and
Fig.~\ref{fig3} which show that for all initial energies $E_{s}$ simulated 
survival probability does not depend, in fact, on $v_{cut}$ until 90\%
(for $E_{s}$ = 1 TeV) to 99.5\%
(for $E_{s}$ = 10 PeV) muons have been stopped. 

It was shown above 
{\it what is result} of simulations with
different models of ionization and values of $v_{cut}$. It was a special 
point of interest for us to track {\it how does it} influence upon 
behaviour of survival probability.

\begin{figure}
\mbox{\epsfig{file=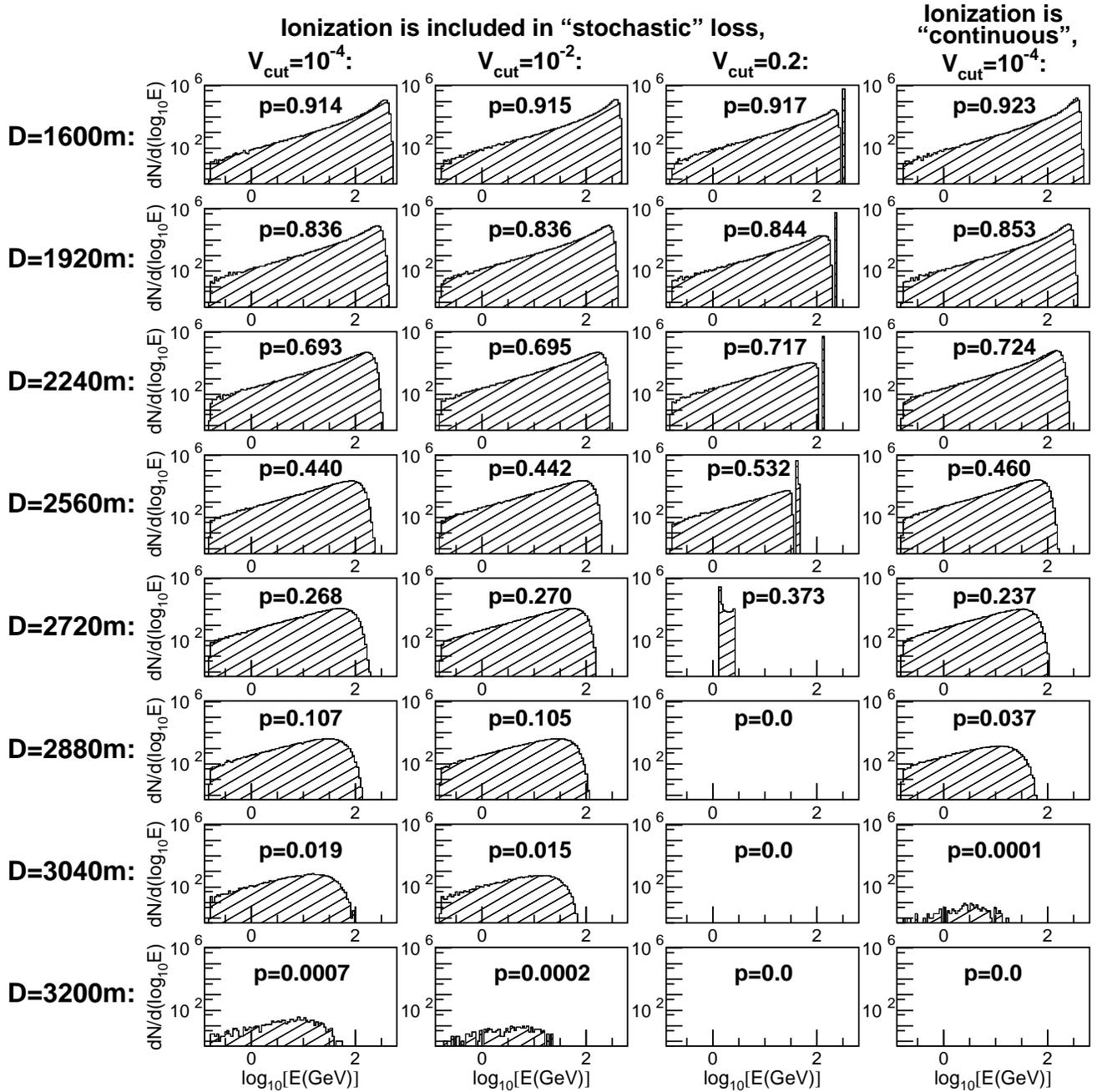,width=17.4cm}}
\caption{Muon spectra at different depths $D$ in pure water resulting from 
10$^{6}$ muons with initial energy $E_{s}$ = 1 TeV obtained by simulation 
with four models. The first three columns represent spectra obtained with 
ionization included in SEL for $v_{cut}$ = 10$^{-4}$ (1st column), 
10$^{-2}$ (2nd column), and 0.2 (3rd column). 4th column contains spectra 
obtained for entirely ``continuous'' ionization and $v_{cut}$ = 10$^{-4}$. 
On each plot value of survival probability $p$ is indicated (without
statistical error which is negligible for the most plots and, on 
desire, can be easily calculated taking into account that $p=N_{D}/10^{6}$, 
where $N_{D}$ is number of muons which have survived after propagation down
to depth $D$).
}
\label{fig4n}
\end{figure}

\noindent
Fig.~\ref{fig4n} shows how muon spectrum resulting from monoenergetic muon 
beam with initial energy $E_{s}$ = 1 TeV transforms when propagating 
through pure 
water down to the depth of 3.2 km. Simulated results for four settings of 
parameters are presented by four columns of plots. The first three columns 
represent spectra obtained with ionization included in SEL for 
$v_{cut}$ = 10$^{-4}$, 10$^{-2}$ and 0.2. 4th column contains spectra 
simulated with entirely ``continuous'' ionization and 
$v_{cut}$ = 10$^{-4}$. The spectra grouped into the first column represent 
some ``ethalon'' because they were simulated with the most accurate tuning
both for $v_{cut}$ and ionization model.
The first three columns demonstrate that ``compactness'' of spectra at the
same depth is the higher the more value of $v_{cut}$ is. Put your attention
to the right edge of spectra which shifts toward low energies when 
$v_{cut}$ increases (it is the most noticeably for $v_{cut}$ = 0.2). The 
reason is that at any depth energy of the most energetic muons in simulated
beam is determined by CEL. These muons due to statistical fluctuations did
not 
undergo interactions with $v \ge v_{cut}$ and, consequently, lost energy 
only by CEL which increases when $v_{cut}$ increases. That 
is why the maximum
energy in simulated muon beam is lower for large values of $v_{cut}$. 
Fraction of muons which did not undergo an ``catastrophic'' act with 
$v \ge v_{cut}$ till given depth grows with increase of $v_{cut}$ because 
free path between two sequential interactions with $v \ge v_{cut}$ grows 
approximately as $\bar L \propto v_{cut}$. It leads, in particular, to 
distinctly visible picks in spectra for $v_{cut}$ = 0.2 consisted just of 
muons which lost energy only by CEL. Also, some deficit of low energy muons
appears if one sets $v_{cut}$ to a large value. In this case left edge of 
spectrum is provided only with muons which interacted with large fraction 
of energy lost while for smaller $v_{cut}$ an additional fraction of muons 
comes here. As a result simulated spectrum of initial monoenergetic muons 
at given depth is more narrow if $v_{cut}$  is large and, on the contrary, 
more wide if $v_{cut}$ is small. 

Now it is easy to understand how value of $v_{cut}$ influences on simulated 
survival probabilities. When simulated muon beam goes through medium loosing
energy both in CEL and SEL processes, its spectrum is constantly shifting 
to the left (energy decreases). For $v_{cut}$ = 10$^{-4}$ the left part of 
spectrum reaches $E$ = 0 at a smaller depth comparing with
larger $v_{cut}$ 
(because in this case spectrum is wider) and survival probability starts to
decrease. At the same depth survival 
probability for $v_{cut}$ = 10$^{-2}$ 
and $v_{cut}$ = 0.2 is still equal to 1. Thus, for the first part of path 
the 
survival probability is always larger for large $v_{cut}$. At some depth 
(which is equal to approximately 2.8 km for considered case) compactness of
spectra simulated with large $v_{cut}$ starts to play an opposite role. Due
to this compactness and higher CEL muons stop faster comparing with accurate
simulation. So, at the final part of the beam path simulated survival 
probability for large $v_{cut}$ decreases faster comparing with accurate 
simulation and, for instance, for $v_{cut}$ = 0.02 the rest of muon beam 
which reaches the depth of $D$ = 2.72 km (37\%
of initial number of muons) completely vanishes within the next 30 m of 
path, while some fraction of muons simulated with $v_{cut}$ = 10$^{-4}$ 
(0.07\%)
escapes down to the depth of $D$ = 3.2 km.

Qualitatively the same effect leads to the same results if one treats
ionization as completely ``continuous'' energy loss. Again, spectra becomes
more narrow since fluctuations in ionization do not work and, as a 
consequence,  survival probability becomes significantly higher comparing 
with simulation with accurate treatment of ionization at the beginning of 
muon beam path and falls down essentially faster at the final part of path. 

Results presented in Sec. \ref{sec:mono} show the significant influence 
which both model of ionization and value of $v_{cut}$ have over survival 
probability for monoenergetic muon beam. But for practical purposes the 
more important is {\it how this factors do work for real atmospheric muons 
with a power spectrum}? 

\subsection{Propagation of muons sampled according to a power sea-level spectrum} 
\label{sec:spectrum}

\begin{figure}
\hspace{1.5cm}
\mbox{\epsfig{file=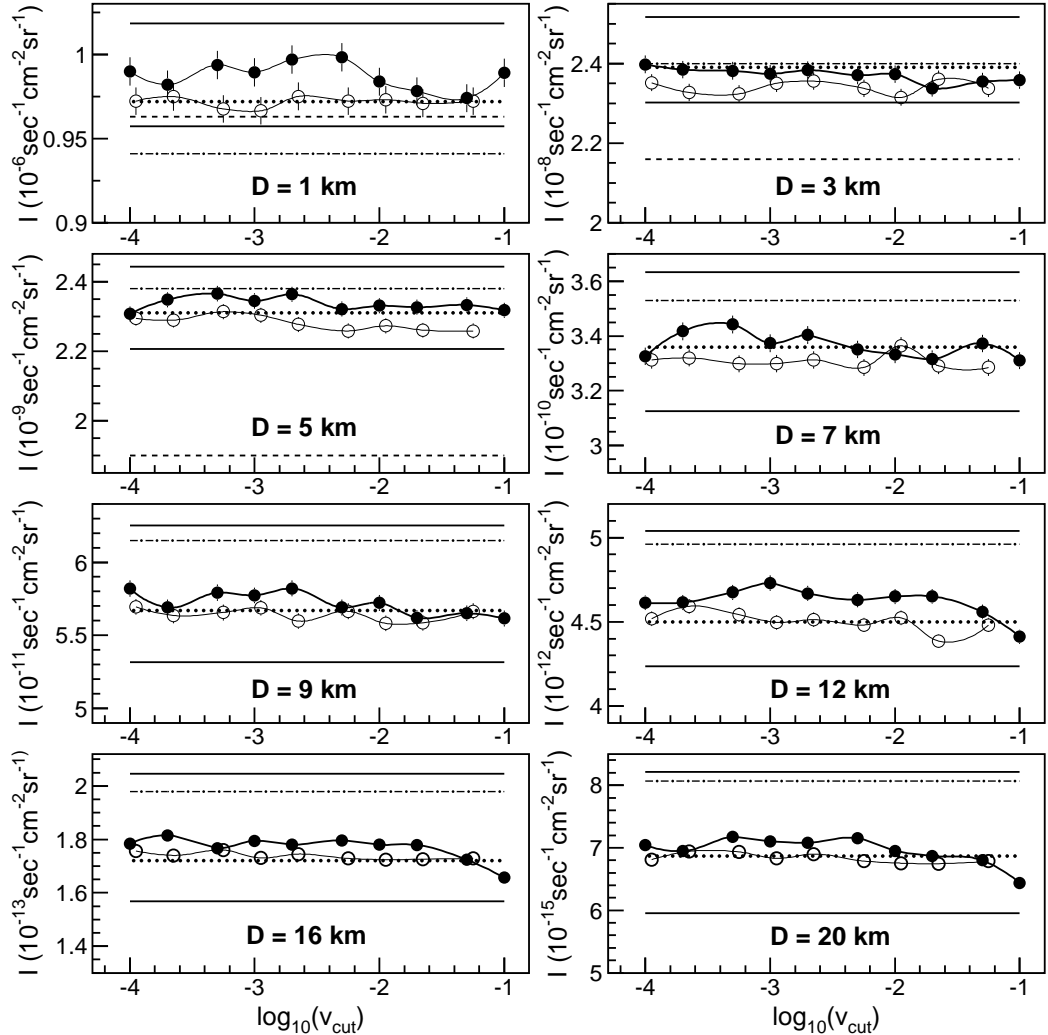,width=13.8cm}}
\protect\caption{
Intensity of vertical atmospheric muon flux $I$ at different depths $D$ 
of pure water vs. $v_{cut}$ as obtained by simulation with muons sampled 
according to sea-level spectrum from Ref.~\protect\cite{bks1} 
(Eq.~(\protect\ref{bknsspec})). Closed circles: ionization is included in 
SEL;
open circles: ionization is completely ``continuous''. Two horizontal solid
lines on each plot show value for survival probability simulated with all
muon cross sections multiplied by a factor  $k_{\sigma}$ = 1.01 (lower 
line) and $k_{\sigma}$ = 0.99 (upper line) for $v_{cut}$ = 10$^{-4}$. 
Dashed lines on plots for $D \le$ 5 km corespond to intensity which was
calculated for all energy loss treated as ``continuous''. Dash-dotted lines
show intensity of vertical muon flux simulated with ionization included in
SEL, $v_{cut}$ = 10$^{-4}$ and muons sampled according to Gaisser sea level
spectrum (Ref. \protect\cite{gaisser}, Eq.~(\protect\ref{gaisspec})). 
Horizontal dotted lines correspond to $v_{cut}$ = 10$^{-4}$ and cross 
section for absorbtion of a real photon at photonuclear interaction 
parametrized according to Ref. \protect\cite{ZEUS} 
(Eq.~(\protect\ref{zeuspn})) instead of parametrization proposed in 
Ref.~\protect\cite{bb} (Eq.~(\protect\ref{bbpn})) which is basic in the
MUM code. 
}
\label{fig5n}
\end{figure}

In Fig.~\ref{fig5n} we present intensity of vertical atmospheric muon flux 
$I$ at different depths of pure water $D$ from 1 km to 20 km vs. $v_{cut}$ 
as simulated with muons sampled according to sea-level spectrum 
(\ref{bknsspec}). Simulation continued until 10$^{4}$ muons reached given 
depth. Curves for two models of ionization are shown for each depth along 
with results for $k_{\sigma}$ = 1.00 $\pm$ 0.01 at $v_{cut}$ = 10$^{-4}$, 
parametrization (\ref{zeuspn}) at $v_{cut}$ = 10$^{-4}$, sea-level muon 
spectrum (\ref{gaisspec}) at $v_{cut}$ = 10$^{-4}$ and all energy loss 
treated entirely as CEL (for depths $D \le$ 5 km only).

General appropriatenesses for real muon spectrum are qualitatively the same
as observed for monoenergetic muon beams.

\begin{itemize}
\item[(a)] For all depths at which neutrino telescopes are located it was 
found to be better to take into account fluctuations of energy loss
simulated by {\it any} model than to treat energy loss as completely 
``continuous'': muon flux intensity computed with non-stochastical model 
of energy loss is always less comparing with stochastical model, the 
difference reaches 10 \%  
at 3 km w.e. and 40\%
at 5 km w.e. (at the depth of 20 km of pure water vertical muon flux 
computed with ignorance of fluctuations is only 10 \% 
of simulated flux).
\item[(b)] Like in a case for monoenergetic beams 1\%-uncertainty in muon
cross sections plays the principal role for resulting error in simulated
muon depth intensity. This error has a tendency to grow with depth
from $\pm$2.5\% at depth of 1 km w.e. to $\pm$15\% at 20 km w.e.. 
But a particular case of this uncertainty, namely difference between
parametrizations for total cross section for absorbtion of a real photon by 
a nucleon at photonuclear interaction from Refs.~\cite{bb,ZEUS}, does not
lead to a significant difference in resulting intensity.
\item[(c)] Difference between muon spectra (\ref{bknsspec}) and 
(\ref{gaisspec}) leads to uncertainty from -4\% 
($D$ = 1 km) to 16\% 
($D$ = 20 km). 
\item[(d)] Error which appears due to simplified, entirely ``continuous''
ionization lies, commonly, at the level of 2--3 \%.
\item[(e)] Dependence of simulated muon flux intensity upon $v_{cut}$ is the
most weak one comparing with other studied error sources. Function 
$I(v_{cut})$ is almost a constant if $v_{cut} \le$ 0.02--0.05 and changes 
in a range $\pm$1--2\%
which is very close to statistical error. 
\end{itemize}

\begin{figure}
\hspace{3.2cm}
\mbox{\epsfig{file=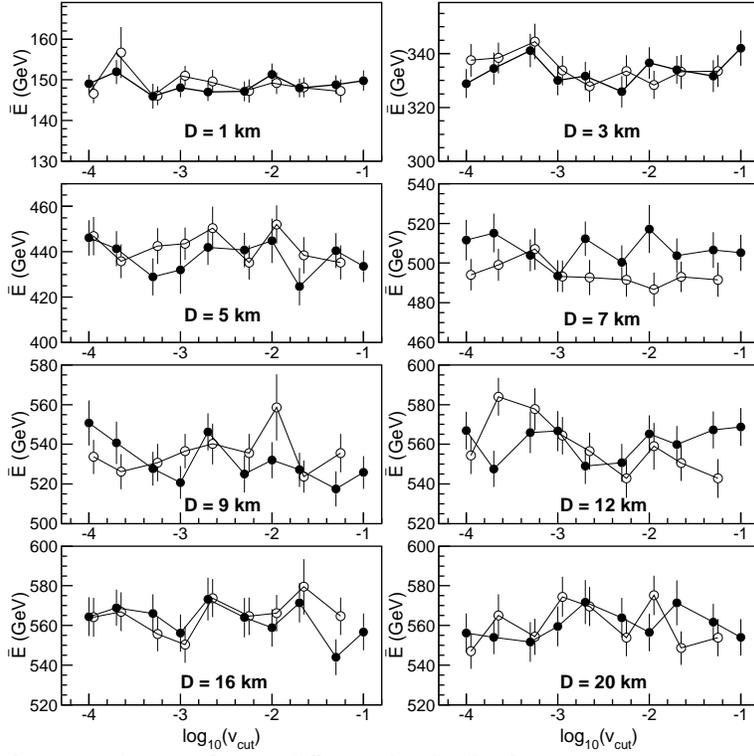,width=10.0cm}}
\protect\caption{
Mean energy for vertical muon flux at different depths $D$ of pure water 
vs. $v_{cut}$ as obtained by simulation with muons sampled according to 
sea-level spectrum from Ref.~\protect\cite{bks1} 
(Eq.~(\protect\ref{bknsspec})). Closed circles: ionization is included in 
SEL; open circles: ionization is completely ``continuous''.
}
\label{fig6n}
\end{figure}

\begin{figure}
\hspace{3.2cm}
\mbox{\epsfig{file=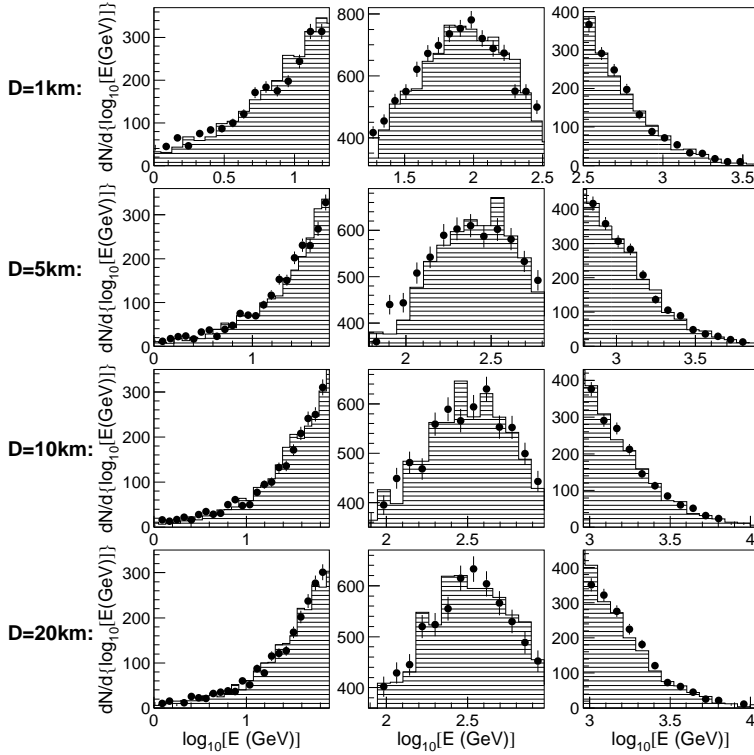,width=10.0cm}}
\protect\caption{
Simulated differential muon spectra at four depths $D$ = 1 km, 5 km, 10 km,
20 km of pure water. Two spectra are displayed on each plot, namely 
simulated for $v_{cut}$ = 10$^{-4}$ (histogram) and $v_{cut}$ = 0.2 (closed
circles). Spectra are normalized for 10$^{4}$ muons with energies $E>$ 0.16 
GeV and at each depth are divided into three parts along muon energy to 
keep linear scale at Y-axis.
}
\label{fig7n}
\end{figure}

We also tried to reveal how value of $v_{cut}$ and model for ionization
influence upon differential muon depth spectra. 
No differences were detected which 
would exceed statistical error. It is illustrated by two figures.
Fig.~\ref{fig6n} displays simulated mean energies for vertical muon flux at
different depths $D$ of pure water vs. $v_{cut}$ as simulated with muons 
sampled according to sea-level spectrum (\ref{bknsspec}). Two dependencies 
are presented at each depth for two model of ionization loss. No 
appropriatenesses are visible on the  plot. In Fig.~\ref{fig7n} we 
present simulated differential muon spectra at four depths $D$ = 1 km, 
5 km, 10 km, 20 km. Two spectra are displayed on each plot obtained 
{\it i)} for the most accurate value of $v_{cut}$ = 10$^{-4}$ and 
{\it ii)} for the most rough case $v_{cut}$ = 0.2. Spectra are normalized 
to 10$^{4}$ muons and at each depth are divided into three parts along 
muon energy to keep linear scale at Y-axis. Again, no statistically 
significant differences are marked.

\begin{figure}
\hspace{5.2cm}
\mbox{\epsfig{file=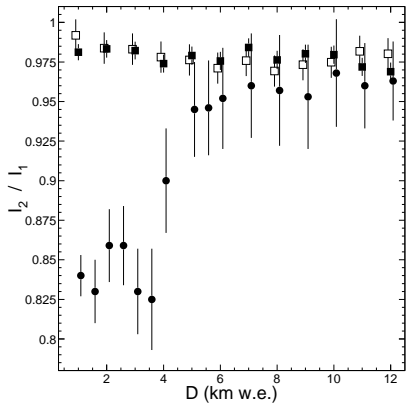,width=5.3cm}}
\protect\caption{
Dependencies for relation $I_{2}/I_{1}$ vs. water equivalent depth in 
standard rock as computed in this work (closed squares), 
in Ref.~\protect\cite{music1} (open squares) and in 
Ref.~\protect\cite{lagutin1} 
(closed circles). $I_{1}$ is depth intensity for vertical muon flux 
simulated with ionization included in SEL, $I_{2}$ is one simulated with
entirely ``continuous'' ionization. Data for this work are obtained
for sea-level spectrum from Ref.~\protect\cite{bks1} 
(Eq.~(\protect\ref{bknsspec})) and $v_{cut}$ = 10$^{-3}$; data from
Ref.~\protect\cite{music1} represent result of simulation for spectrum
from Ref.~\protect\cite{gaisser} (Eq.~(\protect\ref{gaisspec})) and 
$v_{cut}$ = 10$^{-3}$; data from Ref.~\protect\cite{lagutin1} were
simulated with spectrum from Ref.~\protect\cite{volkova} with
``small  transfer grouping'' technique.
}
\label{fig9n}
\end{figure}

\begin{figure}
\hspace{5.2cm}
\mbox{\epsfig{file=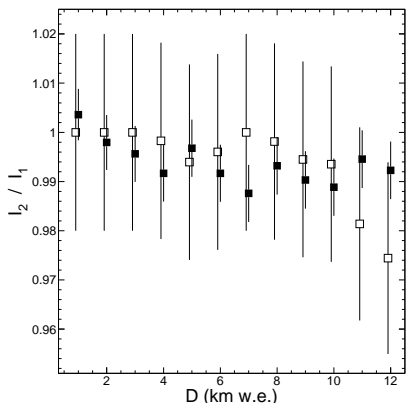,width=5.3cm}}
\protect\caption{
Dependencies for relation $I_{2}/I_{1}$ vs. water equivalent depth in 
standard rock as computed in this work (closed squares) and in 
Ref.~\protect\cite{music1} (open squares). $I_{1}$ is depth intensity for 
vertical muon flux simulated with entirely ``continuous'' ionization and
$v_{cut}$ = 10$^{-3}$, $I_{2}$ is one simulated with the same treatment of
ionization and $v_{cut}$ = 10$^{-2}$. Data for this work are obtained
for sea-level spectrum from Ref.~\protect\cite{bks1} 
(Eq.~(\protect\ref{bknsspec})); data from Ref.~\protect\cite{music1} 
represent result of simulation for spectrum from 
Ref.~\protect\cite{gaisser} (Eq.~(\protect\ref{gaisspec})).
}
\label{fig10n}
\end{figure}

\section{Discussion}
\label{sec:discussion}

Results obtained in this work are evidence of accuracy in parametrizations 
for muon cross sections and spectra to be the principal source of 
uncertainties when simulating muon flux at depths where neutrino 
telescopes are located. Both for monoenergetic muons and muons sampled 
according to a power spectrum it gives uncertainty at least of 2--4\% 
to 10-15\% 
in resulting intensity of muon flux, depending upon depth. Unfortunately, 
this level has at present to be considered as a limit for accuracy of muon 
propagation codes. Influence of model for ionization exceeds this limit 
only for monoenergetic muons with sea-level muon energies $E \le$ 10 TeV 
and only if level of observation is at very last stage of muon path where 
major fraction of initial muon energy has been lost. Actually, due to steep
shape of atmospheric muon power spectrum, an essential part of muons 
reaches detector location being just on the last part of path. Therefore 
effect remains noticeable also for real atmospheric muons but in this case 
uncertainty was found in this work to  be much less: 2--3\%,
which is in an excellent agreement with Refs.~\cite{N94,music1}, while
Ref.~\cite{lagutin1} predicts much more significant difference (up to 17\%).
We suppose this disagreement may result from the fact that ``small transfer
grouping'' technique used for simulation in Ref.~\cite{lagutin1} 
treats muon cross sections to be constant between two interactions
in contrast with algorithm used in the MUM code. As was 
shown in Ref.~\cite{MUM}, in this case switching off 
the fluctuations in
ionization leads to an additinal amplification in effective energy loss
for muon energies $E\le$ 1 TeV and, consequently, simulated muon flux 
intensity must decrease at relatively small depths where muon spectrum is 
formed by muons with sea-level energies just in a range 
$E_{s} \sim$ 10$^{2}$--10$^{3}$ GeV. In Ref.~\cite{music1} the same 
simplification was used but reported reults were obtained by simulation with
$v_{cut}$ = 10$^{-3}$. At this value of $v_{cut}$ role of correct treatment
for free path is not significant. Choice of 
value for $v_{cut}$ is of even less importance and again, it is more 
critical if one investigates monoenergetic muon beam but for power spectrum
alteration in $v_{cut}$ within $v_{cut} \le$ 0.02--0.05 leads only to 1--2\%
differences in simulated muon flux intensities which, again, is in a good
agreement with  level of errors reported in Ref.~\cite{music1}.
Differences between muon flux intensities 
simulated for different models of ionization and values of $v_{cut}$, as 
obtained in given work and in Refs.~\cite{lagutin1,music1}, are presented 
in Fig.~\ref{fig9n} and Fig.~\ref{fig10n}.

Since computation time which is neccessary for simulation depends strongly 
upon used model of ionization loss and upon $v_{cut}$, it 
seems to be reasonable for most purposes to set $v_{cut}$ to 
$v_{cut} \approx$ 0.01--0.05. 
Anyway, in this case resulting error will be 
less comparing with error caused by accuracy of parametrization for muon 
cross sections and sea-level muon spectrum, if one treats free path $L$ by 
an accurate method (Ref.~\cite{MUM}). 
Note that 
``inner'' accuracy of the MUM code was  found  to be  worse  than ``outer''
one (which results from uncertainties of muon cross sections) if 
ionization is set to ``continuous'' model and $v_{cut} >$ 10$^{-2}$.
For fluctuated ionization inner accuracy  remains high enough at least
till $v_{cut}$ = 0.05 (Ref.~\cite{MUM}). So, using MUM one should
set  $v_{cut}$ to $v_{cut} \le$ 0.01 if ionization  is entirely continuous
and to $v_{cut} \le$ 0.05 for ionization included in SEL. 
For test runs it is possible 
to set $v_{cut}$ even to larger values. On the other hand for some 
methodical purposes it may be useful to simulate fluctuations in knock-on 
electron production and choose more fine $v_{cut}$, e.g., if one wants to 
exclude an additional error when comparing results of simulations for 
different models of muon sea-level spectrum with each other.
 
It is impossible to consider all particular cases and give some strict
conformity between setting of parameters at muon MC propagation code
and problem to be solved. We had for an object to create a code which 
would allow to change easy the most important parameters and to perform
an analysis which would allow anyone to choose these parameters according
to one's purposes and taste. We hope this object has been achieved with this
article.

\section{Conclusions}
\label{sec:conclusions}

We have presented the detailed investigation for dependence of computed 
depth muon flux upon the most important parameters of muon propagation MC 
algorithm, which was done with the MUM (MUons+Medium) code 
(Ref.~\cite{MUM}). 
Contributions of different simplifications and uncertainties to the 
resulting error have been analysed and ranked. 
We hope that presented work can be useful for further development of 
muon transport algorithms 
which are neccessary for adequate analysis of muon data obtained at
underground/water/ice neutrino detectors.   

\acknowledgments

We would like to express our gratitude to I. Belolaptikov, A. Butkevich,
R. Kokoulin, V. Kudryavzev and V. Naumov for useful discussion and 
essential remarks which were taken into account in the final version
of the text. One of us (I.S.) is also grateful to L. Bezrukov and 
Ch. Spiering for the attention  and support.

\newpage


\end{document}